\documentclass[runningheads]{llncs}
\usepackage{graphicx}
\begin{document}
\title{Building Analytics Pipelines for Querying Big Streams and Data Histories with H-STREAM
\thanks{The authors would like to thank Enrique J. Arriaga-Varela who participated in the implementation of the first version of H-STREAM funded by the Mexican CONACyT {Young Scientists' Support Program}. 
Experiments were done at the Extreme Computing group of the Barcelona Supercomputing Centre. We thank Microsoft for the Azure grant we obtained that enabled the scalability tests we performed. }
}

\subtitle{Extended Version}

\titlerunning{H-STREAM}
\author{Genoveva Vargas-Solar\inst{1}
%\orcidID{0000-1111-2222-3333} 
\and
Javier A. Espinosa-Oviedo\inst{2}
%\orcidID{1111-2222-3333-4444} 
}
\authorrunning{G. Vargas-Solar et al.}
\institute{French Council of Scientific Research (CNRS), LIRIS, Lyon, France \\
\email{genoveva.vargas-solar@liris.cnrs.fr}\\
%\url{http://www.springer.com/gp/computer-science/lncs} 
\and
University of Lyon, ERIC-LAFMIA, Bron, France \\
\email{javier.espinosa-oviedo@univ-lyon2.fr}}

\maketitle              % typeset the header of the contribution
\begin{abstract}
This paper introduces H-STREAM, a big stream/data processing pipelines evaluation engine  that  proposes stream processing operators as micro-services to support the analysis and visualisation of Big Data streams stemming from IoT (Internet of Things) environments.  H-STREAM micro-services combine stream processing and data storage techniques tuned depending on the number of things producing streams, the pace at which they produce them, and the physical computing resources available for processing them online and delivering them to consumers. 
H-STREAM delivers stream processing and visualisation micro-services installed in a cloud environment. Micro-services can be composed for implementing specific stream aggregation analysis pipelines as queries. 
The paper presents an experimental validation using Microsoft Azure as a deployment environment for testing the capacity of H-STREAM for dealing with
velocity and volume challenges in an (i) a neuroscience experiment and (in) a social connectivity analysis scenario running on  IoT farms. 

%adapted to the requirements of experimental analysis stemming in smart cities.  
%The experimental validation mainly focuses on a scenario using Microsoft Azure as a deployment environment showing how H-STREAM addresses velocity and volume challenges in a social connectivity analysis scenario running on an IoT farm. 
\end{abstract}
%
%%****************************************************************
\section{Introduction}
%****************************************************************
The Internet of Things (IoT) is the network of physical devices, vehicles, home appliances, and other items embedded with electronics, software, sensors, actuators, and connectivity, enabling these objects to connect and exchange data. 
When IoT is augmented with sensors and actuators, 
it enables the construction of cyber-physical systems, including smart grids, virtual power plants, smart homes, intelligent transportation and smart cities. These systems are data centred where data are produced at different paces and requires an agile and resources aware management to be consumed, archived and analysed, optimising in particular underlying computing, storage and communication resources. 
  Streams processing calls for research that can intelligently process raw data and perform
  "online" analysis combining  streaming data and historical persistent data, that could been collected by IoT devices and that has been stored for postmortem analysis.
  Existing platforms provide efficient solutions for processing streams with parallel execution backends for example Apache Flink \footnote{\url{https://flink.apache.org}}, Kafka \footnote{\url{https://kafka.apache.org/intro}} and message-based infrastructures like Rabbit MQ \footnote{\url{https://www.rabbitmq.com}}. Programmers rely on these platforms to define stream processing operations that consume "mini"-batches of streams observed through temporal windows. The processing operations can be rather complex and ad-hoc to target application requirements. It is technically challenging for analytics-based applications to analyse data postmortem combining it with streams to perform more agile and online analytics tasks. 
  For example, compare the number of people entering a shopping mall in the last hour with the average number of people entering the shopping mall at the time interval every day during the last month. This task is, of course, possible to compute but requires data management and programming skills for doing it properly.
  
  Current advances in data processing and data analytics have shown that it is possible to propose general operations as functions or operators that can be called, similar to queries within databases applications (e.g. Spark programs). Still, the stream/data processing operations remain embedded within programs, and this approach prevents reutilisation and can imply high maintenance costs. Providing platforms that can let combine processing, and analytics operators and apply them on streams and stored data, in the spirit of AI Microsoft Gallery are still to come.  The challenge is to define an architecture/platform with the right components that can wrap operators, manage data collections and streams and execute pipelines of analytics tasks that can continuously deliver analytics results to be consumed by target applications. 

This paper proposes H-STREAM \footnote{{\url{https://youtu.be/WmTgGHY\_FUY} }}a big stream/data processing pipeline enactment engine. It  provides stream processing micro-services\footnote{Micro-service is a software development technique that structures an application as a collection of loosely coupled services. In a micro-service oriented architecture, services are fine-grained and the protocols are lightweight.} for supporting the analysis and exploration of big streams in IoT environments. H-STREAM combines stream processing and data storage techniques that are tuned depending on the number of things producing streams, the pace at which they produce them, and the physical computing resources available for processing them on-line and delivering them to consumers. H-STREAM deploys big stream operators pipelines, called big stream queries, on message queues (Rabbit MQ) and big data processing platforms (Spark) to provide a powerful execution environment.

The experiments present stream processing micro-services 
composed and installed in a cloud environment for implementing specific stream aggregation queries. 
The paper particularly focuses on two experiments using Microsoft Azure as deployment environment introducing respectively velocity  and volume challenges  of an IoT farm.  

Accordingly, the remainder of the paper is organised as follows. 
Section \ref{sec:related} introduces related work regarding stream processing and data harvesting looking into IoT environments. 
The section discusses some limitations and underlines how our work intends to go a step forward concerning those solutions. 
Section \ref{sec:platform} describes the general architecture of H-STREAM with operators as micro-services that are deployed on high performance underlying infrastructures. It also introduces the micro-services composition language it proposes for defining Big Stream processing pipelines as queries.
Section \ref{sec:micro-service} introduces the core of our contribution, a stream processing micro-service; it describes its architecture and the data processing workflow it defines. 
Section \ref{sec:experimental} describes the experimental scenario that applies series of micro-services for processing streams produced continuously for analysing connection logs to evaluate scale up in terms of the number of things and volume of streams. 
Section \ref{sec:conclusion} concludes the paper.

%****************************************************************
\section{Related work}\label{sec:related}
%*************************************************************
 We analyse stream management systems that emerged to query streams from continuous data providers (e.g. sensors, things). 
Stream processing is the processing of data in motion, or in other words, computing on data directly as it is produced or received.
Big Streams can result from a fine-grained continuous reading of the environment, society, and organisations, natural and social phenomena. Observations are done in different conditions and with different devices, and therefore, Big Streams are raw heterogeneous continuously produced data flows that must be processed for extracting useful information.
The systems that receive and send the data streams and execute the application or analytics logic are called stream processors. The primary responsibilities of a stream processor are to ensure that data flows efficiently and the computation scales and is fault-tolerant.

Many stream processors adopt stateful strategies. Stateful stream processing \cite{cardellini2016elastic,carbone2017state,alaasam2019stateful}  maintains contextual state used to store information derived from the previously-seen events.
Before stream processing, streams were often stored in a database, a file system, or other forms of mass storage. Applications would query the data or compute over the data as needed.

Apache Storm\footnote{\url{https://storm.apache.org}} is a distributed stream processing computation framework that is distributed, fault-tolerant and guarantees data processing. A Storm application is designed as a "topology" in the shape of a directed acyclic graph (DAG) with spouts and bolts acting as the graph vertices. Edges on the graph are named streams and direct data from one node to another. Together, the topology acts as a data transformation pipeline.
 Apache Flink\footnote{\url{https://flink.apache.org/}} is an open-source stateful stream processing framework. Stateful stream processing integrates the database and the event-driven/reactive application or analytics logic into one tightly integrated entity. With Flink, streaming data from many sources can be ingested, processed, and distributed across various nodes.
 Flink can handle graph processing, machine learning, and other complex event processing. 
 Apache Kafka\footnote{\url{https://kafka.apache.org/documentation/streams/}} is an open-source publish and subscribe messaging solution. Services publishing (writing) events to Kafka topics are asynchronously connected to other services consuming (reading) events from Kafka - all in real-time. Kafka Streams lacks point-to-point queues and falls short in terms of analytics. 
 Spring Cloud Data Flow\footnote{\url{https://spring.io/projects/spring-cloud-dataflow}} is a microservice-based streaming and batch processing platform. It provides tools to create data pipelines for common use cases. Spring Cloud Data Flows has an intuitive graphic editor that makes building data pipelines interactive for developers.
Amazon Kinesis Streams\footnote{\url{http://aws.amazon.com/kinesis/data-streams/}} is a  service to collect, process, and analyse streaming data in real-time, designed to get important information needed to make decisions on time.
Cloud Dataflow\footnote{\url{https://cloud.google.com/dataflow}} is a serverless processing platform designed to execute data processing pipelines. It uses the Apache Beam SDK for MapReduce operations and accuracy control for batch and streaming data. Apache Pulsar is a cloud-native, distributed messaging and streaming platform. 
Apache Pulsar\footnote{\url{https://pulsar.apache.org/}} is a  high-performance cloud-native, distributed messaging and streaming platform that provides server-to-server messaging and geo-replication of messages across clusters. 
IBM Streams\footnote{\url{https://www.ibm.com/cloud/streaming-analytics}}  proposes a Streams Processing Language (SPL). It powers a Stream Analytics service that allows to ingest and analyse millions of events per second. Queries can be expressed to retrieve specific data and create filters to refine the data on your dashboard to dive deeper.Source.
 
%--------------------------------------------------------
\noindent
{\em Discussion}
%--------------------------------------------------------
The real-time stream processing engines are developed for specific use cases such as IoT, finance, advertisement, telecommunications, healthcare, etc. 
 Data is collected from some data-generating sources. Much data are of no interest, and they can be filtered and compressed by orders of magnitude \cite{lyman2004much,woo2013s,zikopoulos2011understanding,bigdatasourcebook}.
They rely on distributed processing models, where unbounded data streams are processed.  Stream analytics is often performed after the complete scanning of representative data sets. This strategy is inconvenient in real-time stream processing to process the entire data stream at once. Windowing mechanisms emerged for processing data stream in a predefined topology with a fixed number of operations such as join, aggregate, filter, etc.  The challenge is to define these filters in such a way that they do not discard helpful information.
 
 Data streaming calls for research that can intelligently process raw data to a size that its users can handle while not missing the needle in the haystack. Furthermore, “online” analysis techniques must process streaming data on the fly and combine them with historical data to provide past and current analytics of observed environments.

%****************************************************************
\section{H-STREAM for building querying pipelines for analysing streams}\label{sec:platform}
\begin{figure}[ht!] \centering
\includegraphics[width=0.90\textwidth]{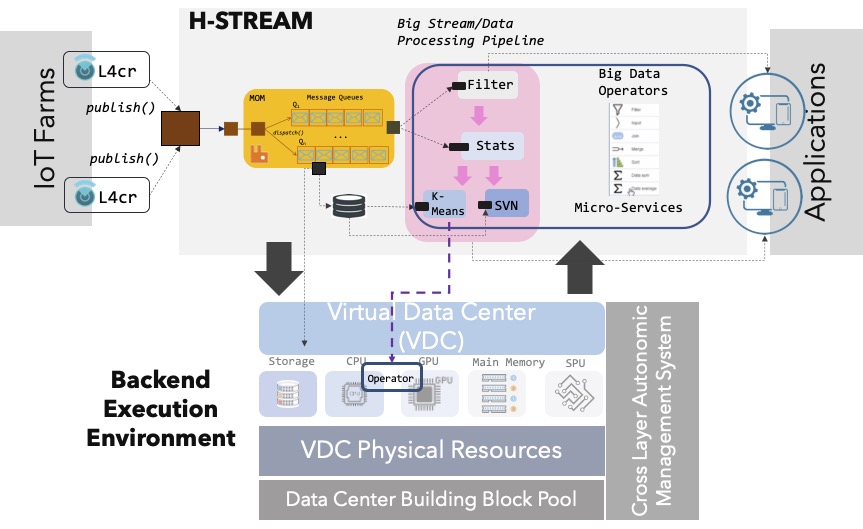}
\caption{H-STREAM General Architecture.}\label{fig:h-stream-architecture}
\end{figure}

%****************************************************************
We propose H-STREAM, a Big stream operators platform implemented as micro-services that can be composed for processing streams stemming from IoT environments (see figure \ref{fig:h-stream-architecture}).  
%A stream is a sequence (a priori infinite) of couples (t$_{i}$, v$_{i}$) where t$_{i}$ is a time stamp and v$_{i}$ is a (simple or complex) value.

H-STREAM is a middleware platform that provides Big Stream/Data operators implementations like aggregation, descriptive statistics, filtering, clustering, and visualisation as micro-services. Operators are composed to define pipelines as queries that apply series of aggregation, filtering and analytics operations to streams stemming from IoT farms and stored data. These pipelines provide analytics visions of data to target applications. H-STREAM relies on (i) message queues for collecting streams online from IoT farms; and (ii) a backend execution environment \cite{akoglu2021putting} that provides a virtual data centre with infrastructure resources necessary for executing costly processes. 

%. . - - . . - -. . - -. . - -. . - -. . - -. . - -. . - -. . - -
\noindent
{\em Composing micro-services}
%. . - - . . - -. . - -. . - -. . - -. . - -. . - -. . - -. . - -
Micro-services can work alone or be composed to implement simple or complex analytics pipelines  (e.g., fetch, sliding window, average, etc.). A series of data processing operations express queries in H-STREAM applied to streams stemming from things, stores or micro-services. A query is implemented by composing micro-services.
The approach for composing micro-services is based on a composition operation that connects them by expressing a data flow (IN/OUT data). We currently compose aggregation services (min, max, mean) with temporal windowing services (landmark, sliding) that receive input data from storage support or a continuous data producer.  We propose connectors, namely {\small\sf Fetch} and {\small\sf Sink} micro-services that determine the way micro-services exchange data from/to things, storage systems, or other micro-services. 
\begin{figure}[ht!] \centering
\includegraphics[width=0.90\textwidth]{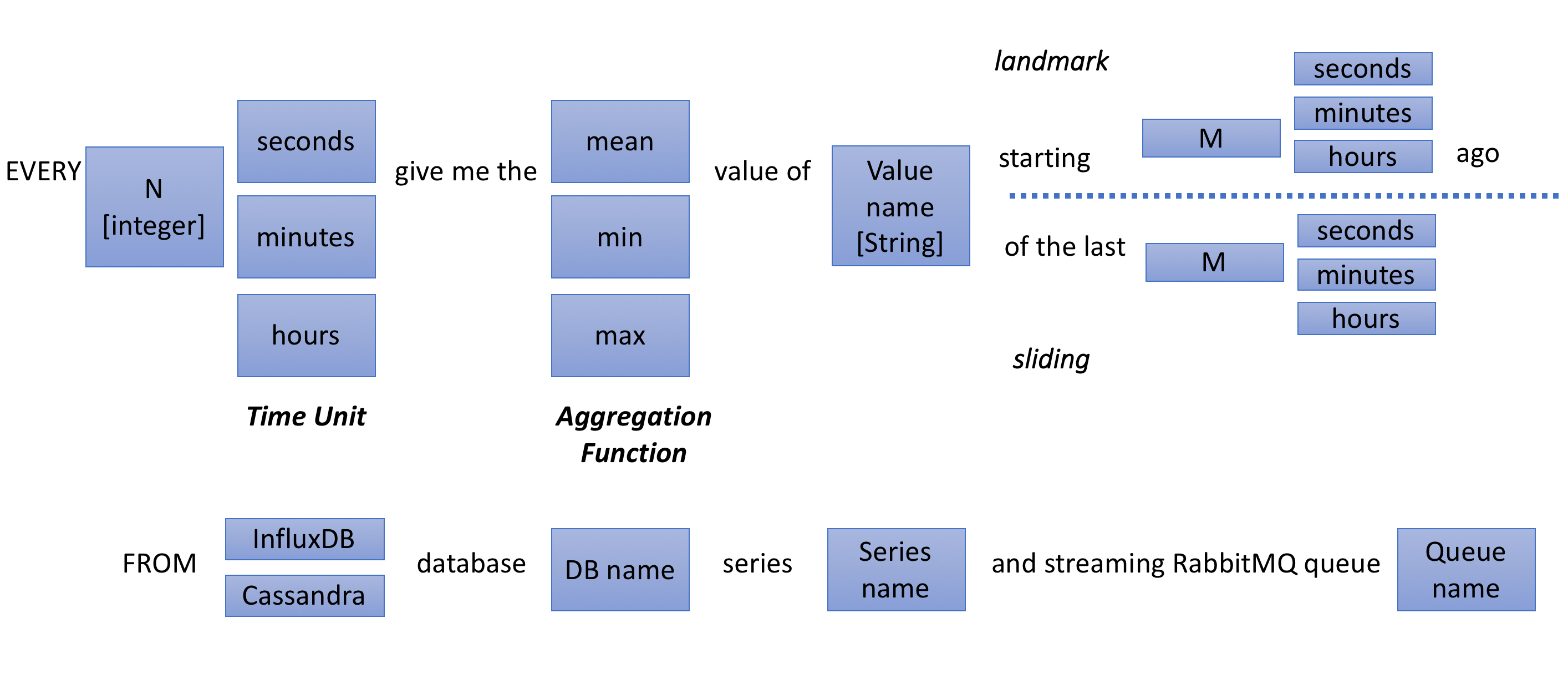}
\caption{Taxonomy of queries that can be processed by composing micro-services.}\label{fig:query-taxo}
\end{figure}

%. . - - . . - -. . - -. . - -. . - -. . - -. . - -. . - -. . - -
\noindent
{\em Big Stream Processing Pipelines Query Language}
%. . - - . . - -. . - -. . - -. . - -. . - -. . - -. . - -. . - -
We proposed a pseudo query language with the syntax 
presented in figure \ref{fig:query-taxo},
 used to express: 
\begin{itemize}
\item The frequency in which data will be consumed ({\small\sf EVERY(number:Integer, timeUnit:\{minutes, seconds, hours\})}).
\item The aggregation function applies to an attribute of the input tuples (min, max, mean).
\item The observation window on top of which aggregation functions will  perform. The window can involve only streams produced online (the last 5 seconds) or include historical data (the last 120 days).
 \end{itemize}
 
The observation can be done starting from a given instance in the past ({\small\sf starting(number:Integer, timeUnit:\{minutes, seconds, hours\})}) until something happens. For example, from the instant in which the execution starts until the consumer is disconnected. This corresponds to a landmark window \cite{golab2010data}. It can also be done continuously starting from a moving "current instant" to several {\small\sf \{minutes, seconds, hours\}} before. This corresponds to a sliding window.
%\end{itemize}
The expression includes the logic names of the data producers that can be a store ({\small\sf  Influx, Cassandra}) and/or a stream queue provided by a message-oriented middleware (e.g., RabbitMQ).

%****************************************************************
\section{Stream processing micro-service}\label{sec:micro-service}
\begin{figure}[ht!] \centering
\includegraphics[width=0.80\textwidth]{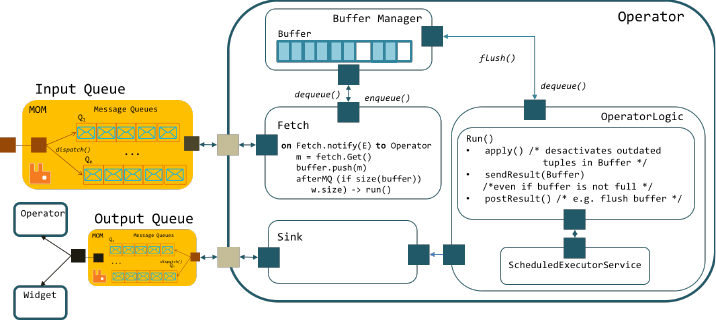}
\caption{Architecture of a stream processing micro-service for processing data streams.}\label{fig:micro-service}
\end{figure}

%****************************************************************
Figure \ref{fig:micro-service} shows the general architecture of a stream micro-service. 
A micro-service consists of three main components, {\small\sf Buffer Manager, Fetch} and {\small\sf Sink}, and {\small\sf OperatorLogic}.  The micro-service logic is based on a scheduler that ensures the recurrence rate in which the analytics operation implemented by the micro-service is executed. Stream processing is based on “unlimited” consumption of data ensured by the component {\sf\small Fetch} that works if a producer notifies streams. This specification is contained in the logic of the components {\small\sf OperatorLogic} and {\small\sf Fetch}.

As shown in the figure, a micro-service communicates asynchronously with other micro-services using a message-oriented middleware. As data is produced, the micro-service fetches and copies the data to an internal buffer. Then, depending on its logic, it applies a processing algorithm and sends it to the micro-services connected to it. 
The micro-services adopt the tuple oriented data model as a stream exchange model among the IoT environment producing streams and the micro-services. A stream is a series of attribute-value couples where values are of atomic types (integer, string, char, float) from a micro-service point of view. 
The general architecture of a micro-service is specialised in concrete micro-services processing streams using well-known window-based stream processing strategies: tumbling, sliding and landmark \cite{kramer2009semantics,golab2010data}. Micro-services can also combine stream histories with continuous flows of streams of the same type (the average number of connections to the Internet by Bob of the last month until the next hour). 

Since RAM assigned to a micro-service might be limited, and in consequence, its buffer, every microservice implements a data management strategy by collaborating with the communication middleware to exploit buffer space, avoiding losing data and generating results on time. A micro-service communicates asynchronously with other micro-services using a message-oriented middleware. As data is produced, the micro-service fetches and copies the data to an internal buffer. Then, depending on its logic, it applies a processing algorithm and sends it to the micro-services connected to it. 
There are two possibilities: 
\begin{itemize}
\item 
(i) on-line processing using tree window-based strategies \cite{kramer2009semantics,golab2010data} (tumbling, sliding and landmark) well known in the stream processing systems domain; (ii)
\item 
combine stream histories with continuous flows of streams of the same type (the average number of connections to the Internet by Bob of the last month until the next hour).
\end{itemize}

\begin{figure}[ht!] \centering
\includegraphics[width=0.90\textwidth]{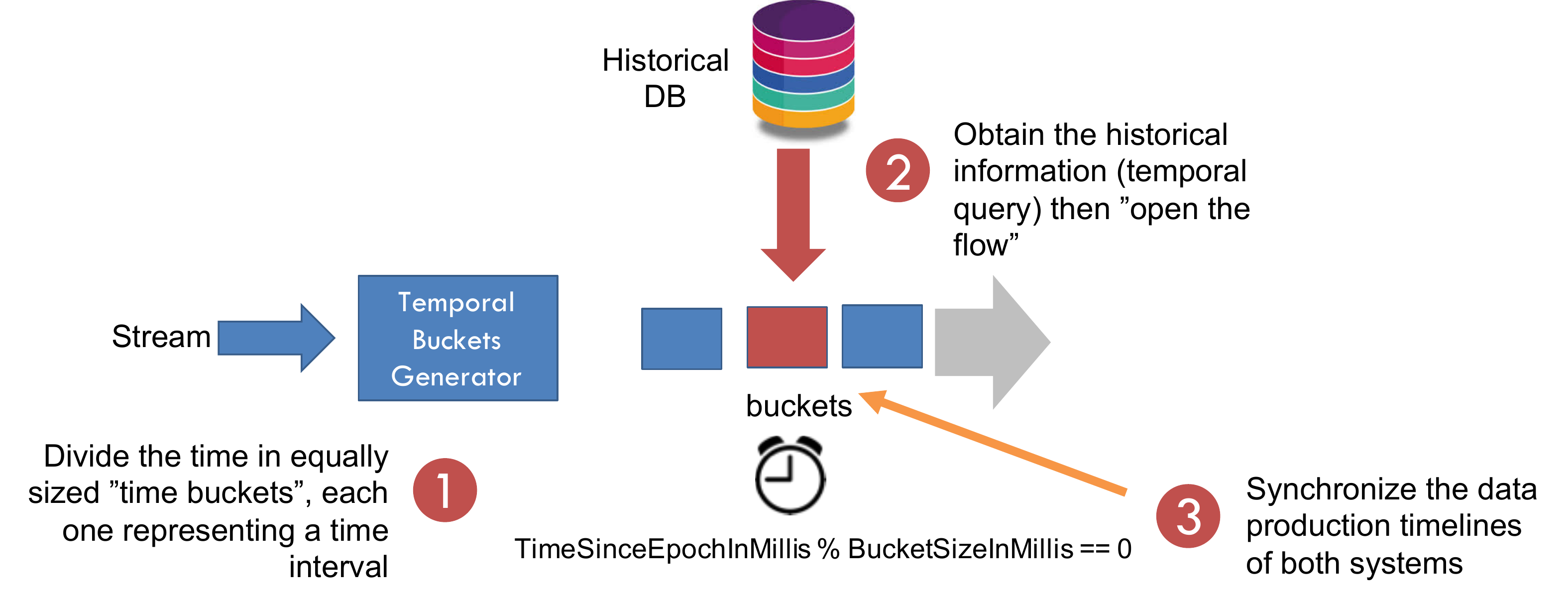}
\caption{Synchronizing stream windows with historic data for computing aggregations.}\label{fig:batch-stream-synchro}
\end{figure}

Figure \ref{fig:batch-stream-synchro} shows the general principle of the functional logic of a micro-service dealing with windows that address the streams harvested before the execution of the current query.
%\end{itemize}	

The following sections describe the strategies adopted for managing memory coupled with cache and local disk to ensure the processing of every tuple consumed by the micro-service; and how to compose micro-services for defining data processing pipelines  used by target applications for querying/analysing data.

%. . - - . . - -. . - -. . - -. . - -. . - -. . - -. . - -. . - -
\subsection{Interval oriented storage support for consuming streams}
%. . - - . . - -. . - -. . - -. . - -. . - -. . - -. . - -. . - -
   A micro-service that aggregates historical data and streams includes a component named {\small\sf HistoricFetch}. This component is responsible for performing a one-shot query for retrieving stored data according to an input query (for example, by a user or application).
 As described above, we have implemented a general/abstract micro-service that contains a {\small\sf Fetch} and {\small\sf Sink} micro-services. 
The historical fetch component has been specialized to interact with two stores:
 \begin{itemize}
 \item 
InfluxDB\footnote{InfluxDB is a time series system accepting temporal queries, useful for computing time tagged tuples (\url{https://www.influxdata.comis})} and
 \item	
Cassandra\footnote{Cassandra is a key-value store that provides non-temporal read/write operations that might be interesting for storing huge quantities of data (\url{http://cassandra.apache.org})}.
 \end{itemize}
The specification of the micro-service {\small\sf  HistoricFetch} and its specialisations are briefly described in the following lines. 
The micro-service {\small\sf HistoricFetch} exports the following interface:\\
\noindent
{\sf\small def queryToHistoric(function: String, value: String,} \\
\hspace*{2,8cm} {\sf\small startTimeInMillis: Long, endTimeInMillis: Long,} \\
\hspace*{2,8cm} {\sf\small             groupByTimeNumber: Int, groupByTimeTimeUnit: String)} :\\
\hspace*{9,8cm} {\sf\small               List[List[Object]] }

The method {\small\sf queryToHistoric()}, shown in the code above, implements the connection to a data store or DBMS, sends queries and retrieves data. 
It returns the results packaged in the Scala structure {\small\sf List[ List[Row] ]} objects, where {\small\sf Row} is a tuple of three elements:
\begin{itemize}
\item {\small\sf Timestamp: Long}, a timestamp in the format epoch; 
\item {\small\sf Count: Double}, the number of tuples (rows) that were grouped;
\item {\small\sf Result: Double}, the result of the aggregation function.
\end{itemize}

As shown in the following code, a component {\small\sf HistoricFetch} is created by the micro-service specifying the name of the store ({\small\sf historicProvider}), the name of the database managing the stream history ({\small\sf dbName.series}) and the execution context. The current version of our micro-service runs on Spark, so the execution context represents a Spark Context ({\small\sf sc}).
{\small
\begin{verbatim}
  val hf : HistoricFetch = new 
  HistoricFetch(historicProvider,dbName,series,sc)
\end{verbatim}
   } 

% Figure \ref{fig:storage-protocol} shows the interaction protocol established with a data store for retrieving historical data. As shown in the figure 

%________________________________________________________________________
%\paragraph{Interaction protocol with data stores.} 
%________________________________________________________________________

The component {\small\sf HistoricFetch} of the micro-service creates an object {\small\sf HistoricProvider} (step 1) that is used as proxy for interacting with a specific store (i.e., InfluxDB or Cassandra). The store synchronously creates an object  {\small\sf Connection} (step 2). The {\small\sf Connection} object will remain open once the query has been executed and results received by {\small\sf HistoricFetch}. Then, the component {\small\sf HistoricalFetch} will use it for sending a temporal query using its method {\small\sf queryToHistoric()} (step 3). The result is then received in the variable {\small\sf Result} that is then processed (i.e., transformed to the internal structure of the operator see below) and shared with the other components through the {\small\sf Buffer} of the micro-service (step 4).

% \begin{figure}[ht!] \centering
% \includegraphics[width=0.95\textwidth]{Figures/storage-protocol.png}
% \caption{General interaction protocol between an micro-service with a Store.}\label{fig:storage-protocol}
% \end{figure}

\begin{figure}[ht!] \centering
\includegraphics[width=0.95\textwidth]{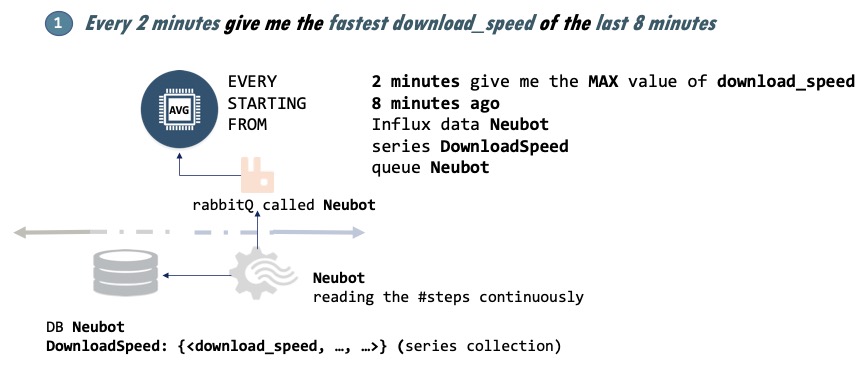}
\caption{Big Streams Query Example.}\label{fig:query-example}
\end{figure}

Consider the observation of download and upload speed variations within users' connections when they work on different networks. Assume that observations are monitored online but that before the query is issued previous observations are also stored. With a window and an average operators it is possible to answer a query {\em every two minutes give me the fastest download speed of the last 8 minutes}. Figure \ref{fig:query-example} shows the corresponding expression in our language and it shows how the logic of the operator deals with streams and data histories. The query asks for a temporal interval filter that includes observations that happened 8 minutes before the moment "current" interpreted as the moment in which the query starts its execution. It also asks to apply  this temporal filter every two seconds. This temporal filtering produces streams/data batches on top of which the operator average is applied.

%We assume that it is possible to navigate through the structure of the tuple for accessing attribute values. We  assume that one of the attributes of the tuple corresponds to its time-stamp. The time-stamp represents the time of arrival of the stream to the communication infrastructure.

%. . - - . . - -. . - -. . - -. . - -. . - -. . - -. . - -. . - -
\subsection{Micro-services execution}
\begin{figure}[ht!] \centering
\includegraphics[width=0.95\textwidth]{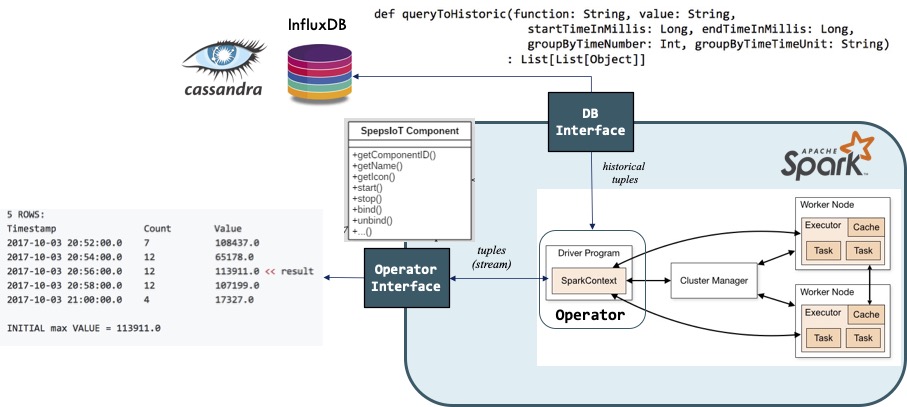}
\caption{Micro-service execution.}\label{fig:micro-service-execution}
\end{figure}

%. . - - . . - -. . - -. . - -. . - -. . - -. . - -. . - -. . - -
Micro-services are executed on top of a Spark infrastructure deployed on a virtual machine provided by the cloud provider Microsoft Azure (see Figure \ref{fig:micro-service-execution}). 
A micro-service running on top of the Spark platform processes streams considering  the following hypothesis:
\begin{itemize}
\item There is a global time model for synchronizing different timelines (batch and stream). 
\item	In the implementation, we use the Spark timeline as a global reference, and we execute aggregations recurrently according to “time buckets”. The size of the time bucket is determined by a query that defines an interval of observation (e.g. the average number of connections of the last 5 hours) and a moving temporal reference (e.g., every 5 minutes, until “now” if “now” is a moving temporal reference). 
\end{itemize}

%. . - - . . - -. . - -. . - -. . - -. . - -. . - -. . - -. . - -
\subsection{Experimental validation}
%. . - - . . - -. . - -. . - -. . - -. . - -. . - -. . - -. . - -
We conducted experiments for validating the use of our micro-services.
The following lines describe the IoT environment that we prepared for our experiments and some scale-up observations. 

For deploying our demonstration experiment, we built an IoT farm using our Azure Grant and implemented a distributed version of the IoT environment to test a clustered version of Rabbit MQ and, therefore, address the scaling-up problem in terms of the number of data producers (things) for our micro-services.

\begin{figure}[ht!] \centering
\includegraphics[width=0.90\textwidth]{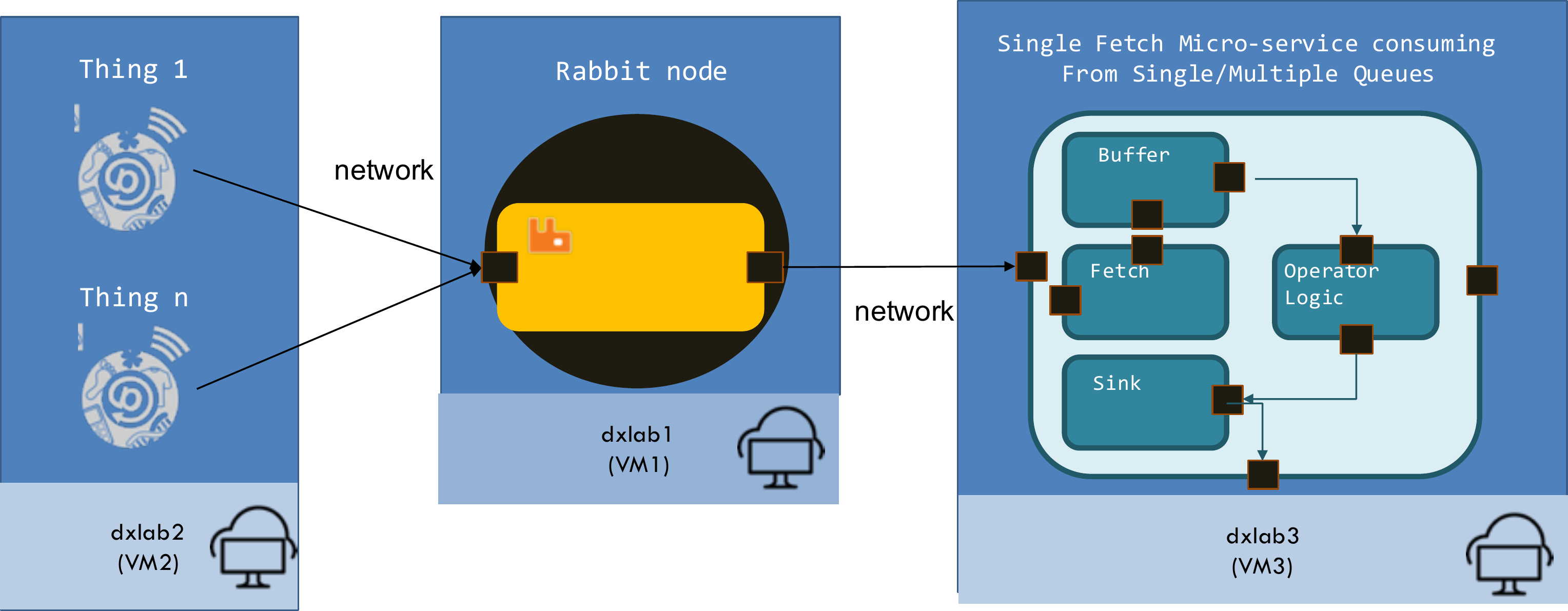}
\caption{General experimental setting deployed on Windows Azure.}\label{fig:rabbit-on-azure-setting}
\end{figure}

Using Azure Virtual Machines (VM) we implemented a  realistic scenario for testing scalability in terms of:
\begin{itemize}
\item	 
(i) Initial MOM (RabbitMQ) installed in the VM$_{2}$ in figure \ref{fig:rabbit-on-azure-setting}.
\item	 
(ii) Producers (Things) installed in the VM$_{1}$ in figure \ref{fig:rabbit-on-azure-setting}.
\item	 
(iii) Micro-services installed in the VM$_{3}$ in figure \ref{fig:rabbit-on-azure-setting}.
\end{itemize}

\begin{figure}[ht!] \centering
\includegraphics[width=0.95\textwidth]{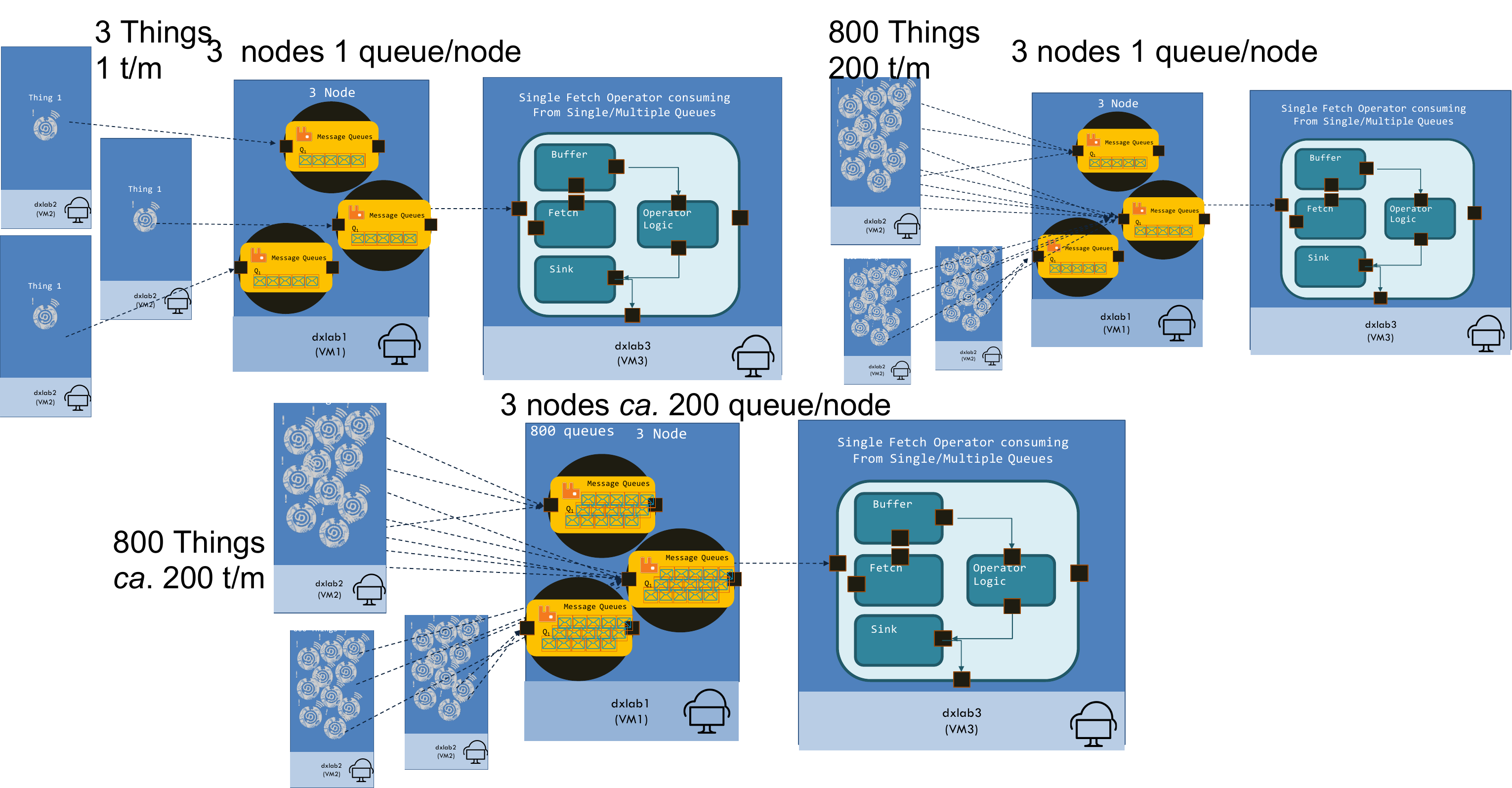}
\caption{Scale up scenarios}\label{fig:rabbit-on-azure-scaleup}
\end{figure}

As shown in figure \ref{fig:rabbit-on-azure-setting}, in this experiment, micro-services and testbed were running on separate VMs. This experiment leads to several cases scaling up to several machines hosting until 800 things with a clustered version of Rabbit using several nodes and queues that could consume millions of messages produced at rates in the order of milliseconds see Figure \ref{fig:rabbit-on-azure-scaleup}).

\begin{figure}[ht!] \centering
\includegraphics[width=0.95\textwidth]{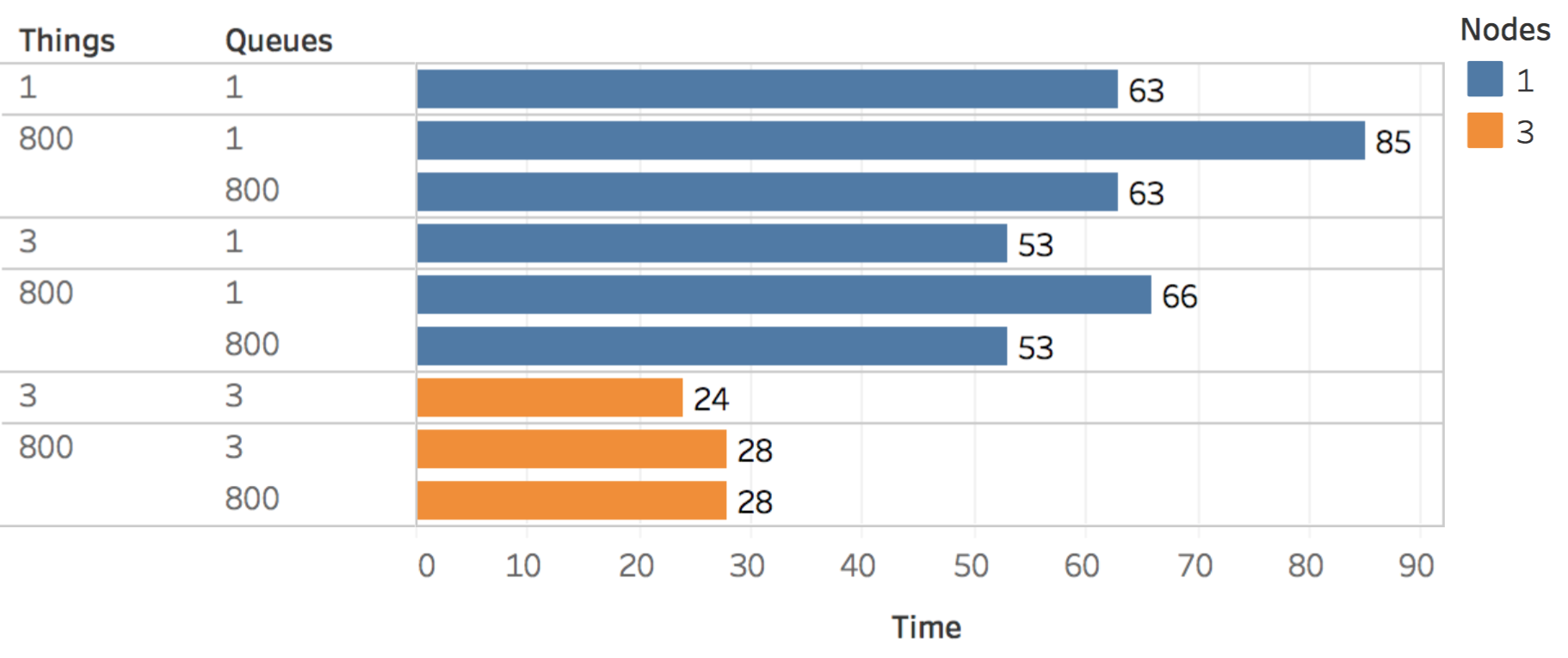}
\caption{Scale up results}\label{fig:rabbit-scaleup}
\end{figure}

 Observations in figure \ref{fig:rabbit-scaleup} showed the behaviour of the IoT environment regarding the message-based communication middleware when the number of things increased, when the production rate varies and when it uses one or several queues for each consuming micro-service. We also observed the behaviour of the IoT environment when several micro-services were consuming and processing the data. Of course, the most agile behaviour is when the number of nodes and virtual machine increases independently of the number of things. Indeed, note that the performance of 800 things against 3 things does not change a lot by increasing nodes, machines and queues. Note also that devoting one queue per thing does not lead to essential changes in performance.
 %
% \vspace{-0,2cm}

For our experimental scenarios, we varied the settings of the IoT environment according to the properties characterising the scenarios. For the first one, regarding the observation of neuroscientific experiments, we used fewer things and queues and more nodes to achieve data processing in an agile way.   Regarding connectivity in cities with many people willing to connect devices in different networks available in different urban spaces, we configured more things and queues and nodes for the second one. 

%https://public.tableau.com/profile/javier.espinosa$#!$/vizhome/Clouding-ThingsRabbitMQ-Cluster/Feuille1

% %****************************************************************
\section{Use Cases}\label{sec:experimental}
% %****************************************************************

The objective of use cases was to have  a proof of concept of the use of our micro-services for implementing stream processing solutions to be weaved in applications. They give
 insight into the way micro-services can be composed to deal with streams production associated with volume and velocity. 
Then we describe two use cases used for experimenting aspect (ii).

%. . - - . . - -. . - -. . - -. . - -. . - -. . - -. . - -. . - -
\subsection{Experimental scenario I: Visual Analytics for Neuroscience}
%. . - - . . - -. . - -. . - -. . - -. . - -. . - -. . - -. .
Neuroscience, like other experimental sciences, supports, refutes or validates hypotheses by conducting experiments on living organisms. For instance, by connecting electrical sensors to a cat’s spinal cord and monitoring its neurons activities, neuroscientists can determine whether capsaicin (chilli pepper active component) has the same effect as anaesthesia in the presence of pain \cite{martin2015machine}. 

\begin{figure}[ht!] \centering
\includegraphics[width=0.95\textwidth]{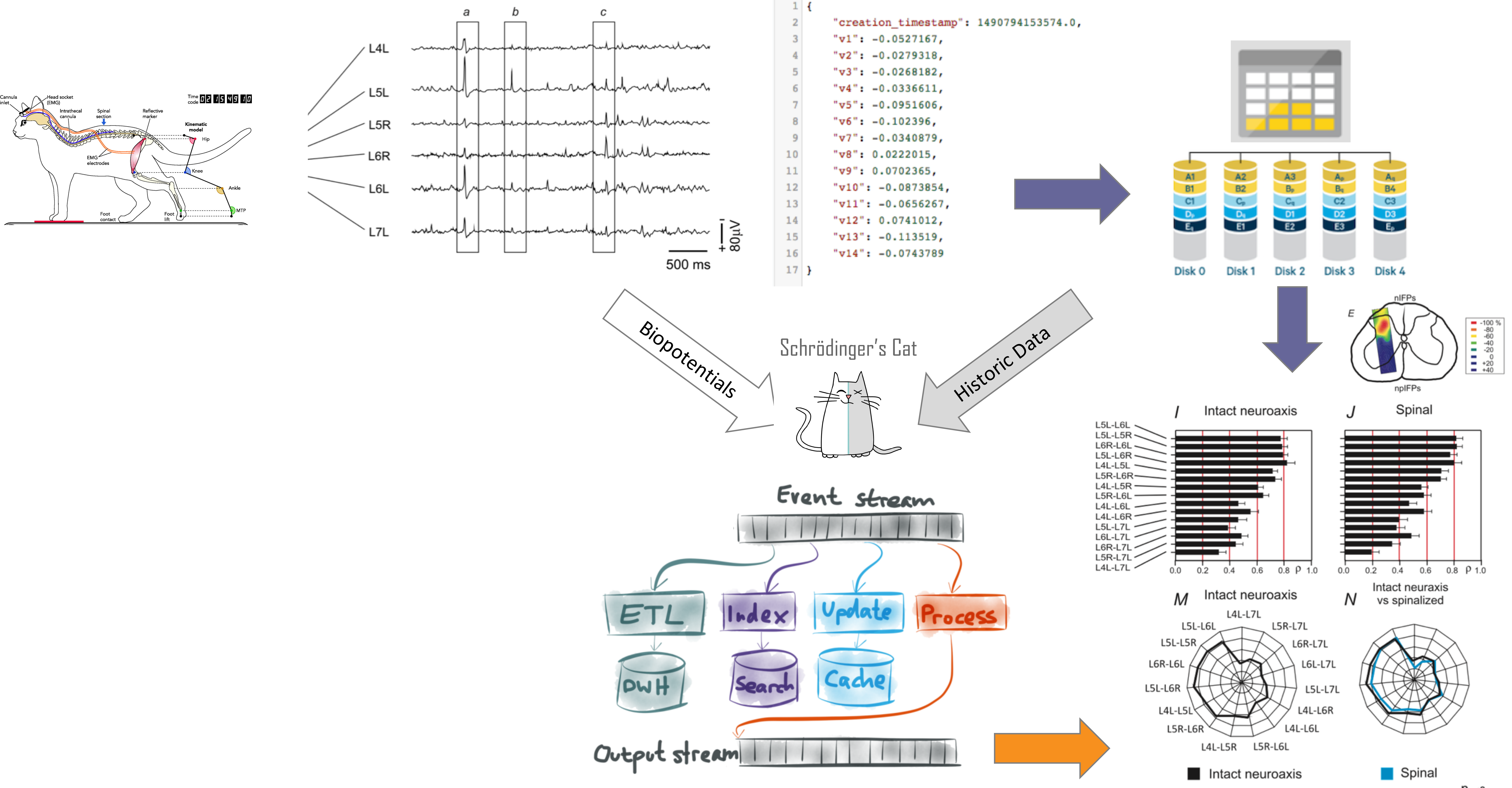}
\caption{Neuroscience scenario.}\label{fig:neuro-scenario}
\end{figure}

In a typical neuroscience experiment 
(see Figure \ref{fig:neuro-scenario}), 
a neuroscientist is responsible for: (i) preparing the subject; (ii) connecting and calibrating sensors; (iii) collecting and storing the experiment data (e.g.file, database); (iv) applying algorithms and statistics for discovering meaningful patterns. Because of the volume of the resulting data and the complexity of the algorithms used for finding patterns, the data analysis is usually done post-mortem. Neuroscientists require novel tools for processing and exploring data in real-time to better control the progress of an experiment. The reason is that the experiment setting is complex (e.g., special and expensive equipment, juridical protocols concerning experiments using animals, gathering together field experts), long (e.g., 8 hours), and unique (e.g., every subject has its proper characteristics).
%Although there has been a lot of progress in the domain of automatic knowledge discovery (e.g.,deep learning), we believe humans play a central role in the data analysis task. 

\begin{figure}[ht!] \centering
\includegraphics[width=0.95\textwidth]{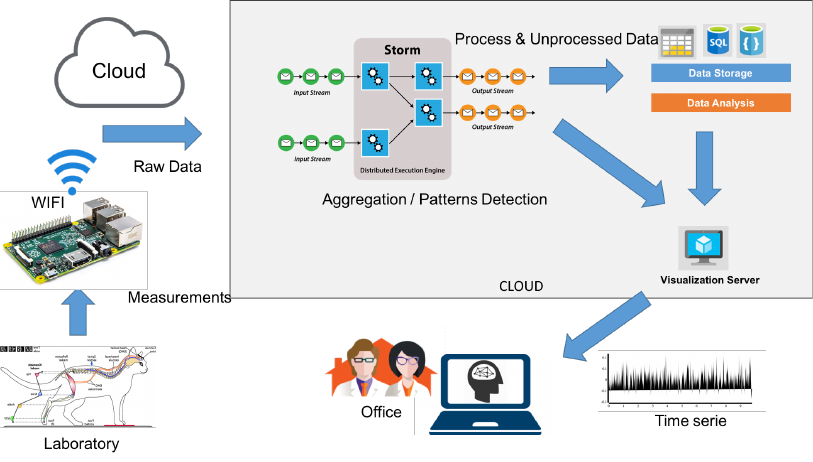}
\caption{Visual stream processing system for analyzing and visualizing data streams.}\label{fig:neuroanalyticsarchitecture}
\end{figure}

Therefore, we provided a solution for harvesting data produced during an experiment and observing specific states of the vertebrae.
Figure \ref{fig:neuroanalyticsarchitecture} gives an overview of our approach. In the figure, data are collected during a neuroscience experiment and continuously transmitted to our system. Data are processed in real-time. Then, depending on the type of analysis that a neuroscientist wants to conduct, she (i) defines queries using a set of operators and (ii) chooses the kind of visual representation that she requires. For instance, in our approach, a neuroscientist can group the data into temporal windows of 1h. Then (for each window), she can choose different visualizations (e.g., point chart, histogram, start plot) for analyzing the correlation among the collected data. 

\begin{figure}[ht!] \centering
\includegraphics[width=0.95\textwidth]{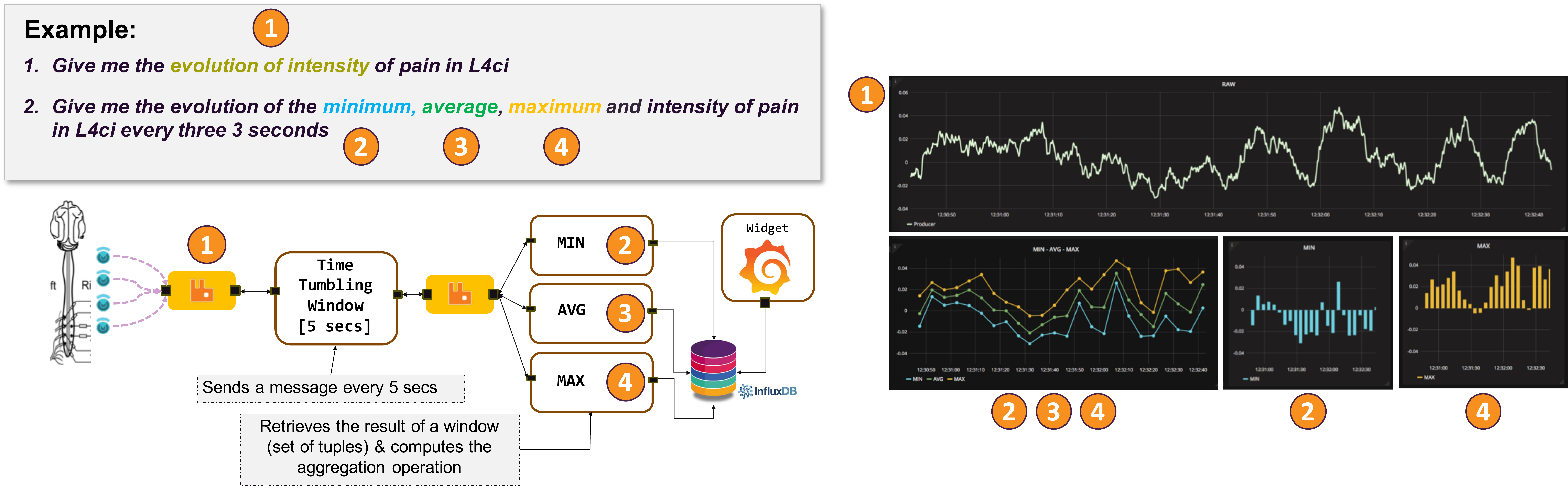}
\caption{Observing continously the execution of the neurscience experiment.}\label{fig:neuro-experiment}
\end{figure}

We have conducted an experimental validation using data from studies regarding pain performed in the neuroscience group at the Mexican research centre CINVESTAV\footnote{Special thanks to Diogenes Chavez from CINVESTAV Department of Physiology, Biophysics and Neuroscience for providing the datasets used in this work.}. 
In this experiment, there is one micro-service for collecting the data from an IoT environment (e.g. an a neuroscience experiment) and one for plotting the data. The idea is that a consumer defines the sequence of micro-services for processing and then plotting the data she desires to observe (as shown in figure \ref{fig:neuro-experiment}). For example, {\em Give me the evolution of pain intensity in L4ci} or {\em Give me the evolution of the minimum, average, maximum and intensity of pain in L4ci every 3 seconds}. Figure \ref{fig:neuro-experiment} shows how these queries are implemented in terms of micro-services, including a window, aggregation, and plotting ones. With this experiment, it was possible to observe online the execution of the neuroscience experiment. Since streams were stored, it was possible to observe the data from other experiments, compare them to make decisions and adjust the phases of the experiment.

%. . - - . . - -. . - -. . - -. . - -. . - -. . - -. . - -. . - -
\subsection{Experimental scenario II: Analysing the behaviour of network services}
%. . - - . . - -. . - -. . - -. . - -. . - -. . - -. . - -. . - -
The experimental scenario  
gives insight on the way micro-services can be composed to deal with volume and velocity. 
For deploying our demonstration experiment, we built an IoT farm using our Azure Grant and implemented a distributed version of the IoT environment to test a clustered version of RabbitMQ and, therefore, address the scaling-up problem in terms of several data producers (things) for our micro-services.

The experimental scenario aims at analyzing the connectivity of the connected society. 
The data set used for this demonstration has been produced in the context of the  Neubot project\footnote{Neubot is a project devoted to measuring Internet from the edges by the Nexa Center for Internet and Society at Politecnico di Torino (https://www.neubot.org/).}.  It consists of network tests (e.g., download/upload speed over HTTP) realized by different users in different locations using an application that measures the network service quality delivered by different Internet connection types
%\footnote{Here the reference of the project that produced the dataset. We omit this information to respect double blinded evaluation requirements.}
\footnote{The HADAS group previously used the Neubot data collection in the context of the FP7 project S2EUNET in collaboration with Politecnico di Torino.}.

\begin{figure}[ht!] \centering
\includegraphics[width=0.90\textwidth]{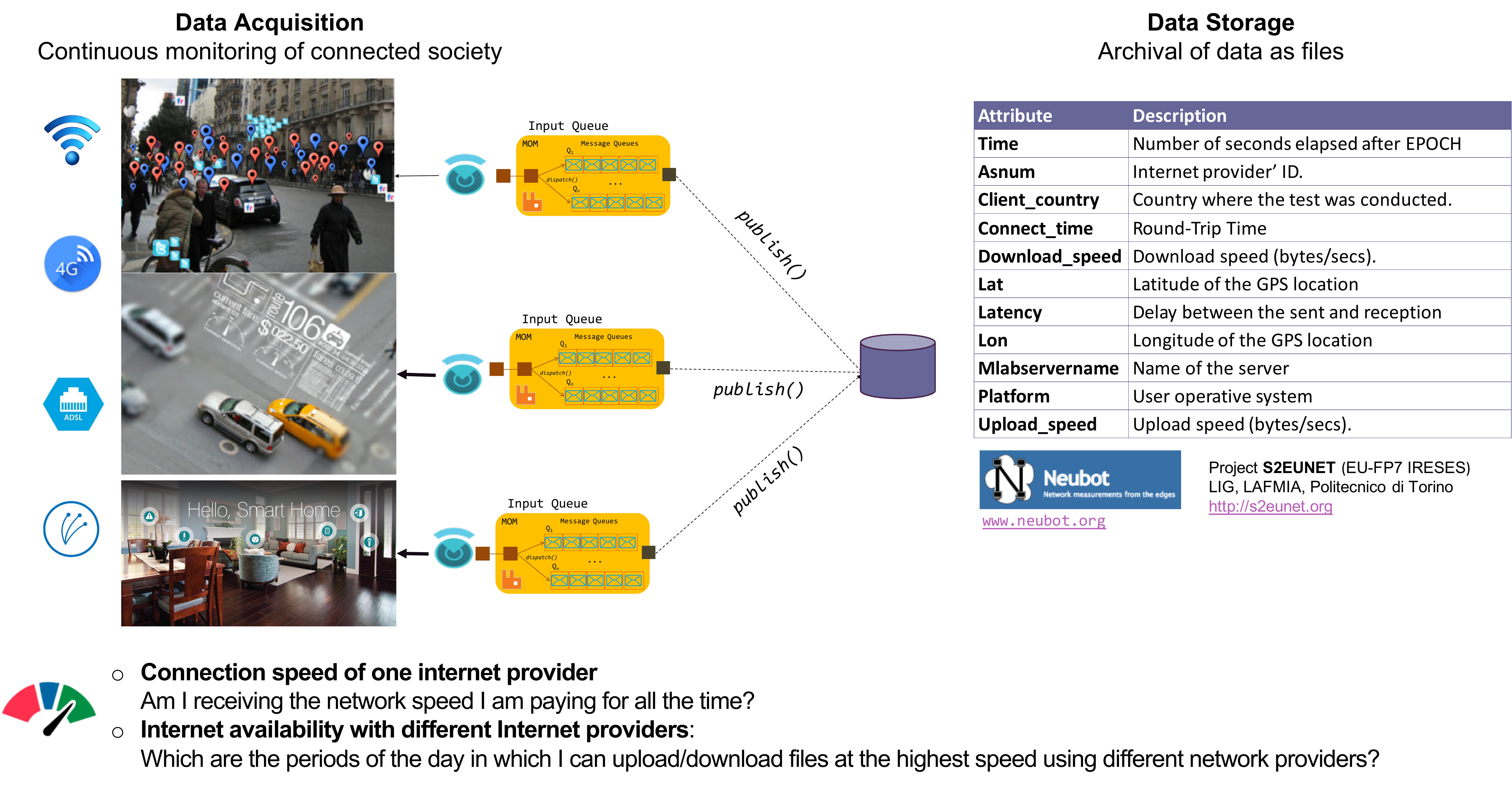}
\caption{Analyzing the connectivity of connected society.}\label{fig:neubot-scenario}
\end{figure}

The idea is that people install the Neubot application on their computers and devices. Every time they connect to the Internet using different networks (4G, Ethernet, etc.), the application computes network quality metrics.
as shown in Figure \ref{fig:neubot-scenario}.\\ 
The data is used then to answer queries such as:
 \begin{itemize}
 \item
The connection speed of one internet provider:
{\em Am I receiving the network speed I am paying for all the time?}\\
\item Internet availability with different Internet providers: 
\noindent
{\em Which are the periods of the day in which I can upload/download files at the highest speed using different network providers?}
 \end{itemize}

\begin{figure}[ht!] \centering
\includegraphics[width=0.85\textwidth]{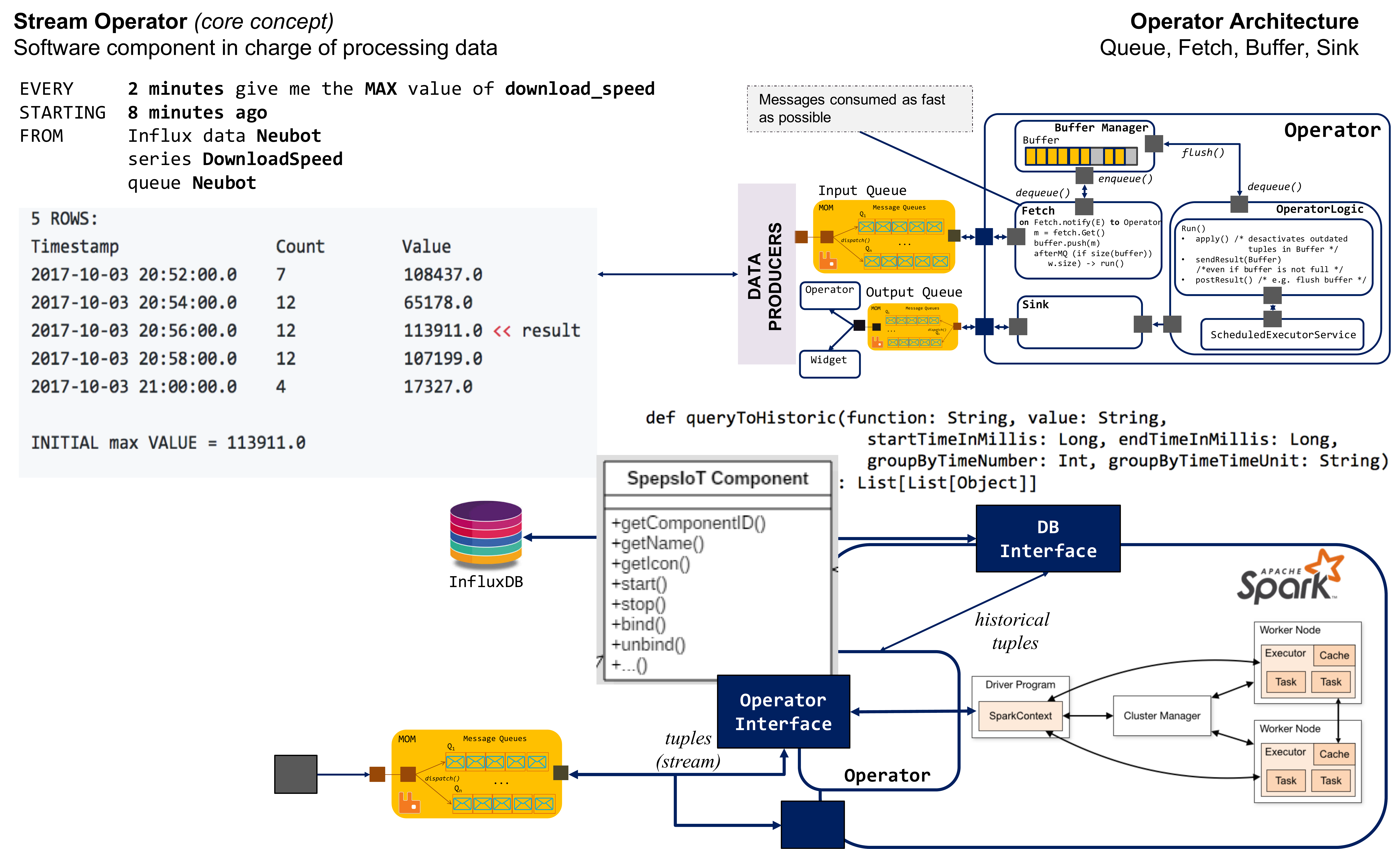}
\caption{Micro-services architecture for analyzing society connectivity with the neubot dataset.}\label{fig:neubot-architecture}
\end{figure}
 
\begin{figure}[ht!] \centering
\includegraphics[width=0.85\textwidth]{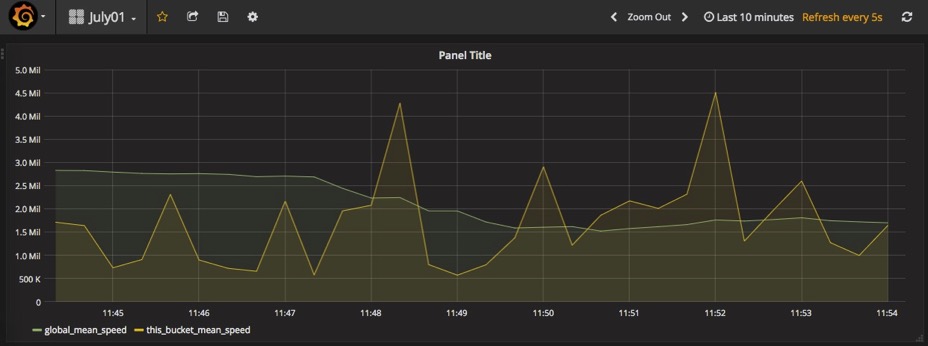}
\caption{H-STREAM query (micro-services composition) output.}
\label{fig:output}
\end{figure}

The type of queries that we tested are the following. The result of an H-STREAM query as a micro-services composition can be seen in Figure \ref{fig:output}:
{\small
\begin{verbatim}
EVERY 20 seconds compute the mean value of download_speed 
      of the last 10 minutes 
FROM influxdb database neubot series speedtest and streaming
     RabbitMQ queue neubotspeed
\end{verbatim}
}

{\small
\begin{verbatim}
EVERY 60 seconds compute the max value of download_speed 
      of the last 3 minutes 
FROM  cassandra database neubot series speedtests and streaming 
      rabbitmq queue neubotspeed

EVERY 	5 minutes compute the mean of the download_speed 
        of the last 120 days 
FROM 	cassandra database neubot series speedtests and streaming 
        rabbitmq queue neubotspeed

EVERY   30 seconds compute the mean value of upload_speed 
        starting 10 days ago 
FROM 	cassandra database neubot series speedtests and streaming 
        rabbitmq queue neubotspeed
\end{verbatim}
}
Depending on the observation window size, the micro-services access the observations stored as post-mortem data sets and connect to online producers currently observing their connections. 
For example, the third query wants to observe a window of 10 days size. 
 Our micro-services could deal with histories produced in windows of size 10 days or even 120 days. Such massive histories could be combined with recent streams and produce reasonable response times (order of seconds).

%****************************************************************
\section{Conclusions and Future Work}\label{sec:conclusion}
%****************************************************************
This paper proposes  H-STREAM  that provides micro-services for analysing and exploring Big  Streams in IoT environments. H-STREAM composes micro-services deployed on high performance computing backends (e.g., cloud, HPC) to process data produced by farms of things producing streams at different paces. 
  Micro-services compositions can correlate online produced streams with post-mortem time series to discover and model phenomena or behaviour patterns of intelligent environments. 
The advantage of this approach is that there is no need for full-fledged data management services; micro-services compositions can tailor data processing functions personalised to the requirements and the characteristics of the applications and IoT environments.

Future work consists of developing a micro-services composition language that can be used for expressing the data processing workflows that can be weaved within target application logics \cite{akoglu2021putting}. We are particularly working on two urban computing projects regarding the modelling and management of crowds %\footnote{\url{https://concernsproject.wordpress.com}}
and smart energy management in urban clusters. 

%  %----------------------------------------------------------------
% \section{Technical requirements}
%  %----------------------------------------------------------------
% The demonstration runs on an environment consisting of a single instance of MoM RabbitMQ\footnote{https://www.rabbitmq.com/} configured with the communication protocol AMQP. We used Apache Spark version 2.1.0 as execution environment and micro-services were programmed using Scala version 2.11.8.
% The demonstration will be presented on the personal computer with a connection to  Microsoft Azure where some services are deployed.

% %----------------------------------------------------------------
% \section{Acknowledgement}
% \begin{footnotesize}
%   The authors would like to thank Enrique Javier Arteaga who participated in the implementation of the first version of the micro-services funded by the CONACYT of the Mexican government {Young Scientists' Support Program}. 
% Javier A. Espinosa-Oviedo and Enrique Javier Arteaga were working at the Extreme Computing group of the Barcelona Supercomputing Centre, where we performed our experiments. We thank Microsoft for the Azure grant we obtained that enabled the scalability tests we performed. 
% \end{footnotesize}
%

 \bibliographystyle{splncs04}
 \bibliography{bibliography}

\end{document}